\documentstyle[aps,twocolumn]{revtex}
\input{psfig}
\begin{document}
\draft
\title{Electron-Electron Interaction in Disordered Mesoscopic Systems:
Weak Localization and Mesoscopic Fluctuations of 
Polarizability and Capacitance}
\author{Ya.~M.~Blanter$^{a}$ and A.~D.~Mirlin$^{b,c}$}
\address{$^a$ D\'epartement de Physique Th\'eorique, Universit\'e de
Gen\`eve, CH-1211 Gen\`eve 4, Switzerland\\
$^b$ Institut f\"ur Theorie der Kondensierten Materie,
Universit\"at Karlsruhe, 76128 Karlsruhe, Germany\\
$^c$ Petersburg Nuclear Physics Institute, 188350 Gatchina,
St. Petersburg, Russia}
\date{\today}
\maketitle 
\tighten
\begin{abstract}
The weak localization correction and the mesoscopic fluctuations of
the polarizability and the capacitance of a small disordered sample 
are studied systematically in  2D and 3D geometries. 
While the grand canonical ensemble calculation gives the
positive magnetopolarizability, in the canonical ensemble (appropriate
for isolated samples) the sign of the effect is reversed. 
The magnitude of mesoscopic fluctuations for a single  sample
exceeds considerably the value of the weak localization correction.

\end{abstract}

\section{Introduction}

The phenomena of weak localization (WL) and mesoscopic fluctuations in
disordered systems have been intensively studied during the last
fifteen years, mainly in connection with transport properties of these
systems \cite{AA,ucf}. For these phenomena the role of the
electron-electron interaction is just in setting the length scale
$l_\phi$ (phase breaking length), below which the electron wave
function preserves its phase coherence. Systems with a size $L$
less than $l_\phi$ are called mesoscopic systems.

In this paper, we consider quantum corrections and mesoscopic
fluctuations of the two other characteristics of a mesoscopic system,
where the electron-electron interaction is essential: polarizability
and capacitance. The former quantity can be measured by putting a
sample into a capacitor, while the latter one determines the charging
energy which shows up in the Coulomb blockade experiments
\cite{Kastner}.  

The first quantum calculation of the polarizability of a small
metallic particle was obtained in a seminal paper by Gor'kov and
Eliashberg (GE) \cite{GE}. It relied on the following
two assumptions concerning statistical properties of energy levels and
eigenfunctions in disordered systems:
\begin{enumerate}
\item The single-particle energy spectrum exhibits the same statistics
as the eigenvalue spectrum of random matrices from the Gaussian
Ensemble of the corresponding symmetry;
\item Exact single-particle eigenfunctions $\psi_k(\bbox{r})$ and
$\psi_l(\bbox{r})$, which are close enough in energy, are correlated
in the following way: 
\end{enumerate}
\begin{equation} \label{func}
V^2 \left\langle \psi^*_k (\bbox{r}) \psi_l (\bbox{r}) \psi_k
(\bbox{r'}) \psi^*_l (\bbox{r'}) \right\rangle_{\epsilon,\omega}
= \Pi_D (\bbox{r}, \bbox{r'}).
\end{equation}
Here the average is defined as
\begin{eqnarray} \label{aver0}
& & \left\langle \psi^*_k (\bbox{r}) \psi_l (\bbox{r}) \psi_k
(\bbox{r'}) \psi^*_l (\bbox{r'}) \right\rangle_{\epsilon,\omega}
\equiv \\
& & \frac{\displaystyle{\langle \sum_{k\ne l} \psi^*_k (\bbox{r})
\psi_l (\bbox{r}) \psi_k (\bbox{r'}) \psi^*_l (\bbox{r'})
\delta(\epsilon - \epsilon_k) \delta(\epsilon + \omega - \epsilon_l)
\rangle}}{\displaystyle{\langle \sum_{k\ne l} \delta(\epsilon -
\epsilon_k) \delta(\epsilon + \omega - \epsilon_l)
\rangle}}, \nonumber   
\end{eqnarray}
$V$ is the sample volume, and the diffusion propagator
$\Pi_D$ is a solution to the diffusion
equation,
\begin{equation} \label{diff}
-D \nabla^2 \Pi_D (\bbox{r}, \bbox{r'}) = (\pi\nu)^{-1} \left[
\delta(\bbox{r - r'}) - V^{-1} \right],
\end{equation}
with the boundary conditions $\nabla_{\bf n} \Pi_D = 0$.
The first of these conjectures was proven by Efetov \cite{Efetov}, and
the second by the authors \cite{BM1}. More precisely, it was shown in
\cite{BM1} that for the energy difference much less than the Thouless
energy, $\omega \ll E_c$, 
\begin{eqnarray} \label{func1}
& & V^2 \left\langle \psi^*_k (\bbox{r}) \psi_l (\bbox{r}) \psi_k
(\bbox{r'}) \psi^*_l (\bbox{r'}) \right\rangle_{\epsilon,\omega}
\nonumber \\
& & = k_d(\bbox{r - r'}) + \Pi_D (\bbox{r}, \bbox{r'}),
\end{eqnarray}
where $k_d(\bbox{r}) = (\pi\nu)^{-2} \langle\mbox{Im}G^R
(\bbox{r})\rangle^2$ is a short-range function ($G^R(\bbox{r})$ being
the retarded Green's function) explicitly given by 
\begin{equation}
 k_d(\bbox{r}) =  \exp (-r/l) \left\{
\begin{array}{ll} 
J_0^2(p_F r), & \ \ 2D\\
(p_Fr)^{-2} \sin^2 p_Fr, & \ \ 3D 
\end{array}
\right..
\label{kd}
\end{equation}
The short-range part $k_d(\bbox{r - r'})$ of the correlation
function (\ref{func1}) was not taken into account by Gor'kov and
Eliashberg, but it would give only a small correction to their
result \cite{foot3}. 

Based on these conjectures, GE concluded that the polarizability for
very low frequencies $\omega \ll \Delta$ ($\Delta = (\nu V)^{-1}$ is the
mean single-particle level spacing; $\nu$ is the density of states per
spin) is enhanced in comparison with the classical value $\alpha_0
\sim V$, the enhancement factor being of order $(\kappa R)^2$, where
\begin{eqnarray} \label{screen0}
\kappa = \left\{ \begin{array}{lr} (8\pi e^2 \nu)^{1/2}, & \mbox{\rm
3D} \\ 4\pi e^2 \nu, & \mbox{\rm 2D} \end{array} \right.
\end{eqnarray}
is the inverse screening radius. Although the original paper \cite{GE}
gave a new insight into the field (called later mesoscopic physics), and
had a substantial impact on the further development of the condensed
matter physics, this result for the polarizability is incorrect for
the following reason. The paper by Gor'kov and Eliashberg does not
take into account the effects of 
screening: they calculate the polarizability in response to the
local field rather than to the external one  \cite{GCE}.
As was found in Refs. \cite{ABB1,ABB2} (see also \cite{Mehlig}), the
screening restores the classical value of the polarizability, thus
reducing the quantum effects to a relatively small
correction. Evaluation of this correction was recently attempted by
Efetov \cite{Ef1}, who combined the non-perturbative calculation of
the polarization function \cite{BM1} with the electron-electron
interactions taken into account in the RPA approximation. Since the
value of the quantum correction depends on the presence (or absence)
of the time reversal symmetry, it was denoted by Efetov as a ``weak
localization correction to polarizability''; we are following this
terminology in the present article. However, he estimated incorrectly
the contribution of the short-range term in Eq.(\ref{func1}), which
made him to conclude that the weak localization correction is
dominated by this term. As we  show below, this is not the case if the
system size exceeds considerably the mean free path. More recently,
Noat, Reulet, and Bouchiat \cite{Noat} presented a perturbative
calculation of the weak localization correction to the polarizability
in a particular geometry of a narrow 2D ring. They
considered both canonical (CE) and grand-canonical (GCE) ensembles,
and concluded that the correction to the polarizability is
parametrically suppressed in the CE. While essentially confirming
their GCE result, we disagree with the above statement concerning the
CE. We show below 
that the effect in the CE is of the same (up to a coefficient of order
one) magnitude than the GCE one, but has an opposite sign.

As was realized by Berkovits and Altshuler \cite{Berk}, fluctuations
in the polarization function lead to mesoscopic fluctuations of
the polarizability of the sample. They considered a specific thin
film geometry and identified the four-diffuson diagrams giving the
leading contribution to the fluctuations. We will follow their
approach when studying the polarizability fluctuations in 2D and 3D
geometries.

Along with the polarizability, we consider another quantity
characterizing a mesoscopic system, the capacitance $C$. It determines
the charging energy $e^2/C$, which manifests itself in the $I-V$
characteristics of a quantum dot in the Coulomb blockade regime. In
particular, the charging energy represents the main contribution to
the threshold voltage in the excitation spectra 
and to the distance between adjacent conductance peaks in the addition
spectra. Statistical properties of the Coulomb blockade $I-V$
characteristics are attracting a great research interest now
\cite{Marcus2,Sivan1,Wharam,Marcus3,Berk1,BMM,Shkl}, which
motivated us  to consider the WL correction and the mesoscopic
fluctuations of the charging energy. In addition, the capacitance
determines the low-frequency behavior of the impedance of mesoscopic
systems \cite{Buettiker,Gopar}. 

Therefore, the purpose of the present paper is to study {\it
systematically} the WL effects and mesoscopic fluctuations of the
polarizability and capacitance in 2D and 3D geometry. Where it is
necessary, we refine results of previous research. We show that the
polarizability and the capacitance can be treated on the same physical
grounds.  We will also find a simple relation between the magnitude of
the WL correction and that of the mesoscopic fluctuations. The
electron-electron interaction is taken into account in 
the RPA approximation, which  works for $\kappa < p_F$, $p_F$ being
the Fermi momentum. We consider the case of low temperature $T \ll
\Delta$ (thus setting $T=0$ in all formulas), 
and study both grand-canonical and canonical ensembles.

\section{Weak localization correction to the polarizability of small
particles}

We consider an isolated disordered metallic particle $\Omega$ (3D or 2D) 
placed into a uniform external frequency-dependent electric field
$\bbox{E}(\omega)$. We assume that the system is diffusive, $l \ll L$,
where $l$ and $L$ are a mean free path and a typical size of a
particle, respectively. In the RPA approximation the potential
distribution $\Phi(\bbox{r})$ and the electron density
$\rho(\bbox{r})$ in the particle obey the Poisson equation
($e=-|e|$ being the electron charge), 
\begin{eqnarray} \label{Poisson}
\Delta \Phi(\bbox{r}) & = & -4\pi e \rho(\bbox{r}) \theta_{\Omega}
(\bbox{r}) \left\{ \begin{array}{lr} 1, & \mbox{\rm 3D} \\
\delta(z), & \mbox{\rm 2D} \end{array} \right. , \nonumber \\ 
& &  \theta_{\Omega} (\bbox{r}) = \left\{ \begin{array}{lr} 1, &
\bbox{r} \in \Omega \\ 0, & \mbox{\rm otherwise} \end{array} \right. ,
\end{eqnarray}
in combination with the equation
\begin{equation} \label{polar1}
\rho(\bbox{r}) = -2e \int_{\Omega} \Pi(\bbox{r},\bbox{r'})
\Phi(\bbox{r'}) d\bbox{r'}.
\end{equation}
In 2D we use the following convention throughout the paper: $\bbox{r}
= (x,y)$ denotes the coordinates in plain, and $z$ the transverse
coordinate. The Laplacian $\Delta$ is always a three-dimensional
operator, 
$\Delta\equiv\Delta_3 = \Delta_2 + \partial_z^2$. Furthermore, $\Pi$
is the polarization function (per spin), which can be readily
expressed through the Matsubara Green's functions, 
\begin{eqnarray} \label{polar2}
\Pi(\bbox{r},\bbox{r'},\omega) & = & -T \sum_{\epsilon_m} \left\langle
G(\bbox{r},\bbox{r'}, i\epsilon_m + i\omega_n) \right. \nonumber \\
& & \times \left. G(\bbox{r'},\bbox{r}, i\epsilon_m) \right\rangle
\vert_{i\omega_n \to \omega + i0},
\end{eqnarray}
or, in terms of the retarded and advanced Green's functions
$G^{R,A}(\bbox{r}, \bbox{r'}, \epsilon)$,
\begin{eqnarray} 
&&\Pi(\bbox{r},\bbox{r'},\omega)\nonumber\\ && =  \frac{-i}{2\pi } 
\int d\epsilon  \left
\{ n_F(\epsilon) \langle G^{R} 
(\bbox{r}, \bbox{r'}, \epsilon + \omega) G^{R} (\bbox{r'},
\bbox{r}, \epsilon) \rangle - \right. \nonumber \\
&& -  \left. n_F(\epsilon + \omega) \langle G^{A}
(\bbox{r}, \bbox{r'}, \epsilon + \omega) G^{A} (\bbox{r'},
\bbox{r}, \epsilon) \rangle + \right. \nonumber \\
&&+ \left. [n_F(\epsilon + \omega) - n_F(\epsilon)]  \langle G^{R}
(\bbox{r}, \bbox{r'}, \epsilon + \omega) G^{A} (\bbox{r'},
\bbox{r}, \epsilon) \rangle \right\} \nonumber\\ 
&& \equiv  \Pi^{RR}(\bbox{r},\bbox{r'},\omega)+
\Pi^{AA}(\bbox{r},\bbox{r'},\omega)+
\Pi^{RA}(\bbox{r},\bbox{r'},\omega).
\label{polar2a}
\end{eqnarray} 
While $\Pi^{RR}$ and $\Pi^{AA}$ can be calculated in the usual
impurity perturbation theory, the evaluation of $\Pi^{RA}$ 
for low frequencies, $\omega \lesssim \Delta$,
requires a non-perturbative treatment. This was done
in Ref. \cite{BM1} in the framework of the supersymmetric sigma-model
approach, and we present only the results here:
\begin{eqnarray}
&&
\Pi^{RR}(\bbox{r},\bbox{r'},\omega)+\Pi^{AA}(\bbox{r},\bbox{r'},\omega)
\nonumber\\& & ={1\over\pi}\int_{-\infty}^0 d\epsilon\mbox{Im}
\langle G^R(\bbox{r},\bbox{r'},\epsilon)\rangle^2
\nonumber \\ &&+{\omega\over 2\pi i}
[\langle \mbox{Re} G^R(\bbox{r},\bbox{r'},0)\rangle^2
-\langle \mbox{Im} G^R(\bbox{r},\bbox{r'},0)\rangle^2]; \nonumber \\
&& \Pi^{RA}(\bbox{r},\bbox{r'},\omega)=-{\nu\over V} 
-\frac{\omega}{2\pi i} \left\{
\langle \mbox{\rm Re} G^R(\bbox{r},\bbox{r'}, 0) \rangle^2 
\right. \nonumber\\
&& + S(\omega)  \left. \langle \mbox{\rm Im} G^R(\bbox{r},\bbox{r'}, 0)
\rangle^2 + (\pi\nu)^2 (1 + S(\omega)) \Pi_D (\bbox{r},\bbox{r'})
\right\}. \nonumber \\ && \label{polar2b}  
\end{eqnarray} 
Here $S(\omega)$ is a correlation function of the zero-dimensional 
sigma-model, $S(\omega)=-\langle Q_{bb}^{11}Q_{bb}^{22}\rangle$ (in
notations of Ref.\cite{BM1}), given explicitly by
\begin{equation} \label{sw}
S(\omega) = 1 + \frac{2i \Delta^2}{\pi^2 \omega^2} \exp \left(
\frac{\pi i \omega}{\Delta} \right) \sin \frac{\pi\omega}{\Delta}.
\end{equation}
It is related to the two-level correlation function $R_2(\omega)$ as
$R_2(\omega)=\mbox{Re}[1+S(\omega)]/2$. 
Now we decompose the polarization function into
frequency-independent and frequency-dependent parts:
\begin{eqnarray} 
& & \Pi(\bbox{r},\bbox{r'},\omega) = \Pi_0(\bbox{r},\bbox{r'}) +
\Pi_1(\bbox{r},\bbox{r'},\omega), \nonumber \\  
& & \Pi_0(\bbox{r},\bbox{r'}) = \frac{1}{\pi} \mbox{\rm Im}
\int_{-\infty}^0 d\epsilon \langle G^R(\bbox{r},\bbox{r'}, \epsilon)
\rangle^2  - \frac{\nu}{V}, \label{polar3a}
\\ & & \Pi_1(\bbox{r},\bbox{r'},\omega) = -\frac{\omega}{2\pi i} 
 (1 + S(\omega)) \nonumber\\
&&\times \left[
 \langle \mbox{\rm Im} G^R(\bbox{r},\bbox{r'}, 0) \rangle^2 + 
 (\pi\nu)^2 \Pi_D (\bbox{r},\bbox{r'})
\right] \nonumber \\
&& ={\nu\over V}A(\omega)[k_d(\bbox{r} - \bbox{r'}) + \Pi_D (\bbox{r},
\bbox{r'})]\ ,  \label{polar3}
\end{eqnarray} 
where we defined $A(\omega) = (i\pi\omega/2\Delta)(1 + S(\omega))$ and
introduced the function $k_d(\bbox{r} - \bbox{r'})=\langle \mbox{\rm
Im} G^R(\bbox{r},\bbox{r'}, 0)\rangle^2$ given explicitly by
Eq.(\ref{kd}).  Note that the formulas
(\ref{polar2b}), (\ref{sw}), (\ref{polar3}) are written for the case of 
the unitary ensemble (broken time-reversal symmetry due to the
presence of a strong enough magnetic field), which was considered in
Ref.\cite{BM1}. Generalization to the orthogonal ensemble (unbroken
time-reversal symmetry) is straightforward and results in
the following modification of the factor $A(\omega)$ in
Eq.(\ref{polar3})  for $\Pi_1$:
\begin{equation}
A(\omega)= {i\pi\omega\over 2\Delta} (1+S(\omega))-1\ ,
\label{ort1}
\end{equation}
where now
\begin{equation}
S(\omega)=1+{2i e^{is} \sin s \over s^2}+2i{d\over ds}\left({\sin
s\over s}\right)\int_1^\infty {e^{ist}\over t} dt
\label{ort2}
\end{equation}
and $s=\pi\omega/\Delta$. In the limit of low frequency,
$\omega\ll\Delta$, the factor $A(\omega)$ is equal to 
\begin{equation}
A(\omega\ll\Delta)=-{2\over\beta}\equiv\left\{\begin{array}{ll}
-1, & \qquad \mbox{unitary}\\
-2, & \qquad \mbox{orthogonal}
\end{array}\ \ , \right.
\label{aw}
\end{equation}
where $\beta$ is the usual parameter equal to 1 (2) for the orthogonal
(resp. unitary) ensemble. 

Now we turn to the calculation of the dipole moment. The general
expressions were obtained in Ref. \cite{Ef1}; we present here the
derivation for the sake of completeness. We consider the
frequency-dependent part $\Pi_1$ as a perturbation and expand
functions $\rho (\bbox{r})$, $\Phi(\bbox{r})$, and the dipole moment, 
$$\bbox{d} = e \int \bbox{r} \rho (\bbox{r}) d\bbox{r},$$
with respect to $\Pi_1$. In the zero-order approximation we obtain
\begin{equation} \label{dip0}
\bbox{d_0} = - 2e \int_{\Omega} \bbox{r} \Pi_0 (\bbox{r}, \bbox{r'})
\Phi_0 (\bbox{r'}) d\bbox{r} d\bbox{r'} , 
\end{equation}
where the potential $\Phi_0 (\bbox{r})$ satisfies Eqs. (\ref{Poisson}),
(\ref{polar1}) with $\Pi$ replaced  by $\Pi_0$. 

\begin{figure}
\centerline{\psfig{figure=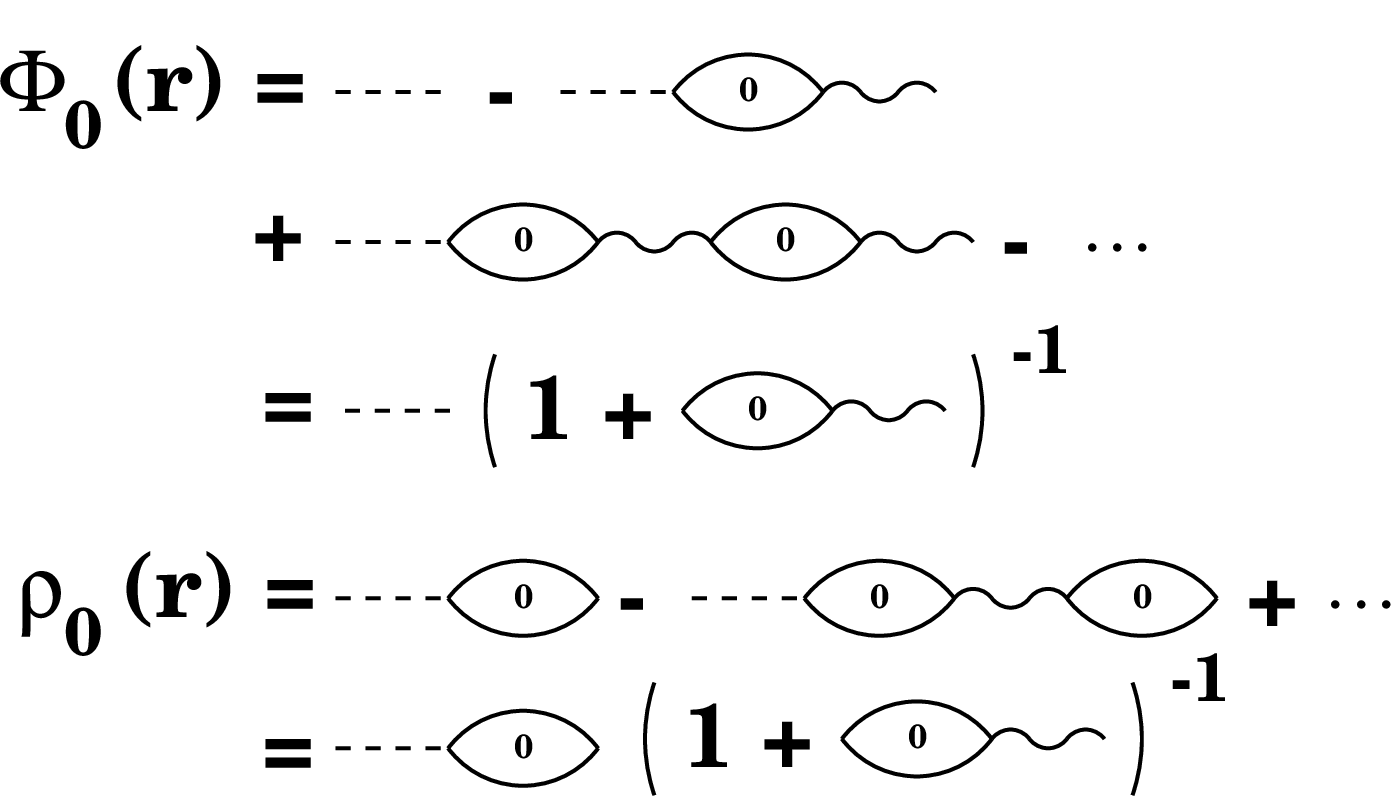,width=7.5cm}}
\centerline{\psfig{figure=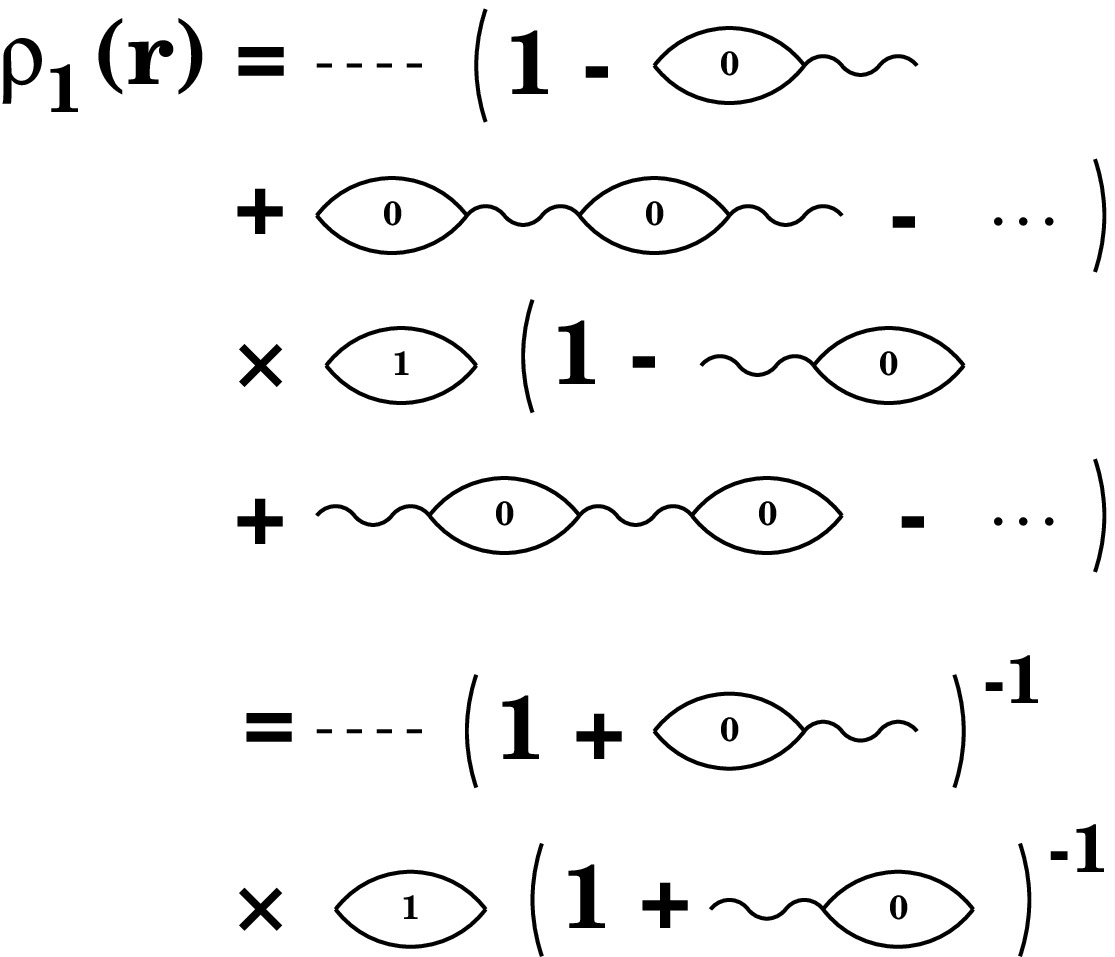,width=6.cm}}
\vspace{0.3cm}
\caption{The electrostatic potential and the electron density in the
RPA approximation. The external dashed line is the bare potential of
the electric field $-{\bf Er}$, the loops with $0$ and $1$ denote the
contributions $\Pi_0$ and $\Pi_1$ to the polarization function,
respectively. The wavy line is the Coulomb interaction.}
\label{fig1}
 
\end{figure}

It is easy to check that the first term in the expression
(\ref{polar3a}) for $\Pi_0(\bbox{r},\bbox{r'})$ gives $\nu$ after
integration over one of the coordinates, the integral being dominated
by the distances of order of the Fermi wave length, 
$|\bbox{r}-\bbox{r'}|\sim\lambda_F$. Assuming the screening length
(which sets the scale for the variation of the potential
$\Phi(\bbox{r})$) to be larger than $\lambda_F$, we can thus replace
this term by a delta-function: 
\begin{equation}
\label{polar4}
\Pi_0(\bbox{r},\bbox{r'}) = \nu \left[ \delta(\bbox{r}-\bbox{r'})
- V^{-1} \right]. 
\end{equation}
This approximation for the polarization function $\Pi_0$ leads to the
simple relation between the potential $\Phi_0(\bbox{r})$ and the excess
density $\rho_0(\bbox{r})$ (Thomas-Fermi approximation)
\begin{equation}
\label{tf}
\Phi_0 (\bbox{r}) = -(2e\nu)^{-1} \rho_0 (\bbox{r}),    
\end{equation}
and consequently to the following equation for the potential 
$\Phi_0(\bbox{r})$,
\begin{eqnarray} \label{phi0}
& & \Delta \Phi_0 (\bbox{r}) = \left\{ \begin{array}{ll} \kappa^2
\Phi_0 (\bbox{r}) \theta_{\Omega} (\bbox{r})\ ,& \qquad 3D \\
\kappa \delta(z) \Phi_0 (\bbox{r}) \theta_{\Omega} (\bbox{r})\ ,&
\qquad 2D\ 
\end{array}\ \ ,  \right.    
\end{eqnarray}
with $\kappa$ defined in Eq.(\ref{screen0}) and
the boundary condition $\Phi_0 (r \to \infty) =
-\bbox{Er}+\mbox{const}$. In Eqs.(\ref{tf}), (\ref{phi0}) we have
chosen the arbitrary additive constant in the definition of $\Phi_0$
in such a way that $\int_\Omega\Phi_0(\bbox{r})d \bbox{r}=0$.

In the following we consider particular
geometries of a 3D spherical sample of a radius $R$ (to be referred as
3D), a circle of a radius $R$ in the in-plane electric field (2D), and
a quasi-two-dimensional sample of a thickness $h$ ($0 < z < h$) and
an area $S$ in the field directed along the $z$-axis,
i.e. transverse to the sample (Q2D) 
The polarizability tensor $\alpha_{ij}$ is generally
defined as $\bbox{d}(\omega) =
\hat\alpha(\omega) \bbox{E}(\omega)$. Eqs. (\ref{dip0}) and
(\ref{phi0}) yield the classical polarizability \cite{LL8} 
equal in the limit $\kappa R\ll 1$, $\kappa h\ll 1$ to 
(for all the geometries under consideration the dipole moment is
directed along the field, and the tensor is reduced to a scalar),
\begin{eqnarray} \label{class1}
\alpha_0 \simeq \left\{ \begin{array}{lr} R^3, & \mbox{\rm 3D} \\
Sh/4\pi, & \mbox{\rm Q2D} \\ (4/3\pi)R^3, & \mbox{\rm 2D}  
\end{array} \right. .
\end{eqnarray} 

Now we turn to the corrections due to the function $\Pi_1$. One
obtains (Fig.1)
\begin{eqnarray} \label{dip1}
\bbox{d_1E} & = & 2e^2 \int d\bbox{r} d\bbox{r'} \Phi_0 (\bbox{r})
\Pi_1 (\bbox{r}, \bbox{r'}) \Phi_0 (\bbox{r'}),
\end{eqnarray}
in full accordance with Ref. \cite{Ef1}. Note that Eq. (\ref{dip1})
depends explicitly  on the symmetry of the system with respect to the
time reversal, and therefore constitutes the WL correction to the
polarizability. As follows from   Eq. (\ref{polar3}), 
this correction consists of two contributions. The first one (to be
referred as $\alpha_{1S}$) is due to the short-range contribution to
the polarization function (the first term in the brackets,
Eq. (\ref{polar3})), and the second one ($\alpha_{1D}$) is due to the
diffusion contribution $\Pi_D$.   

To evaluate  the second term, we use the expansion of $\Pi_D$ in the 
eigenfunctions of the Laplace operator $\phi_{\alpha} (\bbox{r})$ with
the boundary conditions $\nabla \phi_{\alpha} =0$ and the
corresponding eigenvalues $-\epsilon_{\alpha}$,
\begin{equation} \label{expand}
\Pi_D (\bbox{r}, \bbox{r'}) = (\pi D \nu)^{-1} \sum_{\alpha \ne 0}
\epsilon_{\alpha}^{-1} \phi_{\alpha} (\bbox{r}) \phi_{\alpha}
(\bbox{r'}).
\end{equation}
For the purposes of estimate, one can use the following 
expressions valid for $l \ll \vert
\bbox{r} - \bbox{r'} \vert \ll L$
 ($L$ is a typical size of the system), 
\begin{eqnarray} \label{polarest}
& & \Pi_D (\bbox{r},\bbox{r'}) \approx
\left\{ \begin{array}{ll}
\displaystyle{(2\pi^2 \nu D)^{-1} \ln[L /\vert \bbox{r} - \bbox{r'}
\vert]}, & \ 2D\\
\displaystyle{(4\pi^2\nu D \vert \bbox{r} - \bbox{r'}
\vert)^{-1}} , & \ 3D 
\end{array} \right. .
\end{eqnarray}

Now we evaluate and compare both contributions $\alpha_{1S}$ and
$\alpha_{1D}$ in 3D and 2D systems. Since the structure of the
potential $\Phi_0$ is different in 3D and 2D cases, these should be
treated separately. 

\subsection{3D geometry}

For any 3D geometry with $\kappa L \gg 1$ the expression for $\Phi_0
(\bbox{r})$ can be  written in the form 
\begin{equation} \label{appr0}
\Phi_0(\bbox{r}) = \frac{E}{\kappa} \varphi (\bbox{r_{\Vert}})
\exp(-\kappa r_{\perp}), 
\end{equation}
with $\varphi$ being some function of magnitude unity. We have
introduced a transverse coordinate $r_{\perp}$ ($r$ for the sphere,
$z$ in the case of a disk), and the vector $\bbox{r_{\Vert}}$ of
coordinates along the surface of the sample. 
Note that according to Eq.(\ref{tf}),
\begin{equation}
\label{pot3d}
\varphi (\bbox{r_{\Vert}})=-4\pi e\sigma_0(\bbox{r_{\Vert}})/E\ ,
\end{equation}
where $e\sigma_0(\bbox{r_{\Vert}})$ is the  charge density on a
surface of an ideal conductor induced by the electric field $E$. It
can be found by the methods of the classical electrostatics \cite{LL8}.
In the integral for $\alpha_{1S}$,
\begin{eqnarray} \label{Ef3D}
\alpha_{1S} & = & \frac{2e^2\nu A(\omega)}{V E^2} \nonumber \\
& & \times \int d^3\bbox{r}
d^3\bbox{r'} \Phi_0 (\bbox{r}) k_d(|\bbox{r} - \bbox{r'}|)
\Phi_0 (\bbox{r'}),    
\end{eqnarray}
both points $\bbox{r}$ and $\bbox{r'}$ lie in fact in the layer of
thickness $\kappa^{-1} \ll l$ near the surface of the sample. One can
then integrate over the transverse coordinates and reduce the
remaining double surface integral to the integral over one coordinate
only. We obtain 
\begin{eqnarray} \label{polar3Dsur}
\alpha_{1S} & \simeq & \frac{1}{V p_F^2 \kappa^2} A(\omega) \ln
(\kappa l) \int d^2\bbox{r_{\Vert}} \varphi^2(\bbox{r_{\Vert}})
\nonumber \\ & \sim & \frac{1}{L p_F^2 \kappa^2} A(\omega) \ln (\kappa l). 
\end{eqnarray}
In Ref. \cite{Ef1} the kernel in Eq. (\ref{Ef3D}) was incorrectly
replaced by a $\delta$-function, which led to an overestimate of
the contribution $\alpha_{1S}$ 
by a factor of $\kappa l (\ln (\kappa l))^{-1} \gg 1$.   

On the other hand, for the term due to the diffusion,
\begin{eqnarray} \label{diffpol}
\alpha_{1D} & = & \frac{2e^2\nu}{V E^2} A(\omega) \int d\bbox{r}
d\bbox{r'} \Phi_0 (\bbox{r}) \Pi_D (\bbox{r}, \bbox{r'})
\nonumber \\ & & \times \Phi_0 (\bbox{r'}),    
\end{eqnarray}
 we obtain, using the estimate (\ref{polarest}), 
\begin{equation} \label{estim3D}
\alpha_{1D} \sim \frac{1}{\nu D\kappa^{2}} A(\omega) \sim \alpha_{1S} 
\frac{L}{l \ln(\kappa l)}. 
\end{equation} 
We see that in a diffusive system of a size $L\gg l\ln(\kappa l)$, the
diffusion contribution $\alpha_{1D}$ dominates, in contrast to the
conclusion of Ref.\cite{Ef1}. 
At the same time, if the sample size $L$ is comparable to the mean
free path $l$ (which happens e.g. in ballistic systems with surface
scattering), the short-range contribution is parametrically
of the same order (in fact, even larger by a logarithmic factor)
as the diffusive one. 
As expected \cite{ABB1,ABB2,Ef1}, the WL correction is small in
comparison with the classical polarizability $\alpha_0$,  
\begin{equation} \label{estalpha}
\alpha_1 /\alpha_0 \sim {1\over g(\kappa L)^2} A(\omega),$$
\end{equation}
 $g\sim 2\pi\nu DL$ being the dimensionless conductance.
The calculation of the numerical coefficient for the WL correction to
the polarizability requires the exact expansion (\ref{expand}). For
the particular spherical geometry the potential $\Phi_0$ has a form
\begin{equation} \label{classsp}
\Phi_0 (\bbox{r}) = -\frac{3ER}{\pi\kappa r} \exp (-\kappa (R-r)) \cos
\theta, \ \ \ R - r \ll R,
\end{equation}
and we obtain 
\begin{equation} \label{finpol3}
\alpha_1 = {1.36 \over (p_F\kappa)^2 l} A(\omega).
\end{equation}
According to Eq.(\ref{aw}), the  WL correction to the polarizability is
negative. The value of the
correction in the presence of strong magnetic field (unitary symmetry) is
smaller (twice as small for zero frequency) as without the latter
(orthogonal symmetry). The experimentally measured
magnetopolarizability $\alpha_B$, defined as 
\begin{equation} \label{magn}
\alpha_B = \alpha (B) - \alpha(0),
\end{equation}
is therefore positive, in agreement with Ref. \cite{Ef1}.

\subsection{Quasi-2D geometry (transverse field).}
\label{s2b}

We consider now a quasi-two-dimensional sample of a thickness
$h\gg\kappa^{-1}$ and an area $S\gg h^2$ with the electric field
directed transverse to the sample plane. Then Eq.(\ref{appr0}) for the
potential reduces to
\begin{equation}
\label{potq2d}
\Phi_0({\bf r})= {E\over\kappa}(-e^{-\kappa z}+e^{-\kappa(h-z)})
\end{equation}
If the sample is relatively thick, $h\gg l$, the same
consideration as for the case of a spherical shape yields
\begin{equation}
\alpha_{1S}=A(\omega){\ln (\kappa l)\over h (p_F\kappa)^2}\ ,\qquad
\alpha_{1D}={3\over 2}A(\omega){1\over l(p_F\kappa)^2},
\label{q2da}
\end{equation}
and the diffusion term dominates for $h>l\ln(\kappa l)$. In the
opposite case of a thin sample  the short-range contribution is the
leading one. In particular, for $h<l$ we find
\begin{equation}
\label{q2db}
\alpha_1\simeq\alpha_{1S}=A(\omega){\ln (\kappa h)\over h
(p_F\kappa)^2} 
\end{equation}

As is seen from the above formulas, the relative magnitude of the weak
localization correction is rather low for both 3D and quasi-2D (with
the field direction normal to the plane) geometries, so that the
experimental observation of the effect in these cases may be
problematic. The effect is much more pronounced in the 2D case, which
we consider below. 

\subsection{2D geometry (in-plane field).}

In contrast to the 3D case, the potential $\Phi_0$ in the case of a 2D
sample in the in-plane electric field is a smooth
function of coordinates, with the characteristic scale set by the
sample size $R$. Therefore
the kernel in the integral (\ref{Ef3D}), which has a support of order
$l$, can be replaced by a $\delta$-function,
$$(p_Fr)^{-1} \exp(-r/l) \approx 2\pi l p_F^{-1} \delta (\bbox{r}).$$
This gives an estimate 
$$\alpha_{1S} \sim l (p_F\kappa)^{-1} A(\omega) \ln (R/l).$$
On the other hand, 
for the diffusive term (\ref{diffpol}) the estimate (\ref{polarest})
implies
$$\alpha_{1D} \sim R^2 (\kappa g)^{-1} A(\omega) \sim \alpha_{1S}
(R/l)^2 (\ln (R/l))^{-1}.$$
Similarly to the 3D case, the diffusion term $\alpha_{1S}$
dominates for $R\gg l$. The relative magnitude of the quantum
correction can thus be estimated as
$$
\alpha_1/\alpha_0\sim 1/g\kappa R,
$$
with $g=2\pi\nu D=k_F l/2$. 

For the particular case of a circular geometry, the potential $\Phi_0$
is given in the polar coordinates ($r,\theta$) by \cite{LL8}
\begin{equation}
\label{pot2d}
\Phi_0 (\bbox{r}) =  -2E (\pi \kappa)^{-1} r
\cos\theta (R^2 - r^2)^{-1/2}
\end{equation}
An exact calculation gives the value of the quantum correction
$$\alpha_1 (\omega) = 1.53  R^2 (\kappa p_F l)^{-1} A(\omega),$$
and the relative magnitude of the correction is
$$
\alpha_1/\alpha_0= 3.6\times {1\over\kappa p_F l R}A(\omega)
$$

In a recent paper, Noat, Reulet and Bouchiat (NRB) \cite{Noat}
proposed a geometry of a narrow 2D ring (radius $R$, width $W\ll R$)
as more favorable for observation of the effect. In the in-plane
electric field the ring becomes polarized with the one-dimensional
(i.e. integrated over the ring cross-section) charge density 
$$
\rho(\theta)={ER\over e\ln(R/W)}\cos\theta,
$$
and the classical polarizability given by
$$
\alpha_0= {\pi R^3 \over \ln(R/W)}.
$$
Calculating the quantum correction,
we find again that for a diffusive ring $R\gg l$, the contribution 
$\alpha_{1D}$ dominates and gives
\begin{equation}
\alpha_1={R^4\over \nu D W^2\kappa\ln^2(R/W)}A(\omega).
\label{ring1}
\end{equation}
The relative magnitude of the correction is
\begin{equation}
\alpha_1/\alpha_0={1\over\pi g W\kappa \ln(R/W)}  A(\omega),
\label{ring2}
\end{equation}
where $g$ is now the quasi-one-dimensional conductance $g=\nu D W/R$.
These results for the ring geometry are by and large
in agreement with those found  by NRB \cite{Noat}.  Actually,
NRB express the polarizability in terms of the exact eigenfunctions of
electrons, conceptually similarly to the original GE calculation, and
then perform the impurity averaging using the semi-classical
expression for the correlation of the exact single-particle
eigenfunctions (see e.g. \cite{AA,MM}). This calculation yields
correct results for the following reasons. First, the
short-ranged terms in Eq. (\ref{func}), omitted in this calculation,
turn out to be unimportant for the WL correction to the
polarizability. Then, the {\em exact} expression for the long-ranged
(diffusive) part of the eigenfunction correlator (\ref{func})
coincides with the semi-classical result even for $\omega \ll \Delta$,
where the latter generally is not expected to be true. 
This has been proved and discussed previously by the authors in
Ref. \cite{BM1}.  

Similarly, we can consider a quasi-one-dimensional strip of width $W$
and length $L\gg W$ oriented along the electric field direction (which
we choose to be the $z$-axis). Again, the sample polarization is
described by the one-dimensional charge density 
$$
\rho(z)={Ez\over e \ln(L/W)}\ ,
$$
yielding the classical polarizability
$$
\alpha_0={L^3\over 12 \ln(L/W)}\ .
$$
The quantum correction is now equal to
\begin{equation}
\label{q1d1}
\alpha_1\simeq\alpha_{1D}={\pi L^3\over 30 \kappa W g
\ln^2(L/\omega)}A(\omega),
\end{equation}
where $g=2\pi\nu DW/L$ is the dimensionless conductance. We obtain
\begin{equation}
\label{q1d2}
\alpha_1/\alpha_0={2\pi\over 5}{1\over\kappa W g \ln(L/W)}A(\omega)\ .
\end{equation}

Thus, we have found that the WL correction to the 
polarizability can be quite appreciable in 2D (circle) and especially
in quasi-1D (ring or strip) geometries, which gives a possibility of
its experimental observation. These conclusions are in full
agreement with those of NRB, Ref.\cite{Noat}.

\subsection{Canonical ensemble}

The results obtained above were derived for the grand
canonical ensemble, where the chemical potential is fixed by an
external reservoir. In
the Appendix A we present the calculations for the canonical ensemble,
which is more appropriate for the problem in question \cite{foot4}. We show
(Eq. (\ref{CEbas})) that  the CE magnetopolarizability differs
from the GCE result by the coefficient $-2.75$. This means that although
the magnitude of the CE effect is the same  as in the GCE,
 the sign is opposite in the CE case: the magnetic field
suppresses the polarizability.

\section{Weak localization effects in the capacitance}

\subsection{Definitions}

A natural definition of the capacitance in an open system is
\begin{equation} \label{def2}
C_{\mu} = e dQ/d\mu,
\end{equation}
where $Q$ is the total charge of the system. This capacitance 
determining the low-frequency transport properties of the system
was studied thoroughly by B\"uttiker and coworkers 
in Refs. \cite{Buettiker,Stafford} (where it was called
``electrochemical capacitance'').  An
explicit calculation \cite{Buettiker} yields
\begin{equation} \label{dif3}
\frac{e^2}{C_{\mu}} = \frac{e^2}{C_g} + {1\over V\nu(E_F)},
\end{equation}
where $C_g$ is the geometrical capacitance, determined from the
equations of the classical electrostatics \cite{LL8} with 
corrections due to the screening effects, and $\nu(E_F)$ is the
density of states at the Fermi level. The average of the second
term  in the right-hand side of Eq. (\ref{dif3})
is the mean level spacing $\Delta$, which is usually much less than the
average of the first term. However, the fluctuations of the second
term are important \cite{Gopar}.  

Being formally applied to the closed system, Eq. (\ref{def2})
yields infinite fluctuations of the charging energy. Indeed, in
Ref. \cite{Berk1}, where an attempt to calculate the fluctuations of
the compressibility of a closed system has been made, the integral
over energies diverged, and the authors could get a finite result
only by cutting it off at  energies of order $\Delta$. 

A proper generalization of the definition (\ref{def2}) for a closed
system is its ``discrete'' version,
\begin{equation} \label{def4}
e^2/C_{\mu} = \mu(N+1) - \mu(N),
\end{equation}
where $\mu(N)$ is the chemical potential of a {\em closed} system of $N$
electrons. The quantity (\ref{def4}) 
has an important physical meaning: it is equal to the spacing
between two consecutive  peaks in the 
addition spectrum of a quantum dot in
the Coulomb blockade regime. Statistical properties of these spacings
were studied experimentally in Refs.\cite{Sivan1,Wharam,Marcus3} and
theoretically in Ref.\cite{BMM,Shkl}. 

Similarly to the case of an open system (Eq.(\ref{dif3})), the peak
spacing (\ref{def4}) can be decomposed into two parts: level spacing
$\Delta_N$ and (usually much larger) contribution associated with
Coulomb interaction (denoted $E_1$ in Ref.\cite{BMM}). The main
contribution to the latter (and thus to the Coulomb blockade peak
spacing) is given by the charging energy $E_C$ 
defined \cite{BMM} as a constant
part of the effective two-particle interaction potential $U(\bbox{r},
\bbox{r'})$ of the electrons in the sample:
\begin{equation} \label{capdef}
E_C \equiv e^2/C \equiv V^{-2} \int_{\Omega} d\bbox{r} d\bbox{r'}
U(\bbox{r}, \bbox{r'}). 
\end{equation}
In particular, the WL correction to the charging energy 
calculated below is the
dominating term for the WL in the Coulomb blockade peak spacing, and
can be, in principle, measured as the magnetic field dependence of the
peak spacing. In contrast, fluctuations of the charging energy do not
give the dominant contribution to the fluctuations of the peak
spacing, see Sections IV, V, and Ref.\cite{BMM}.

\begin{figure}
\centerline{\psfig{figure=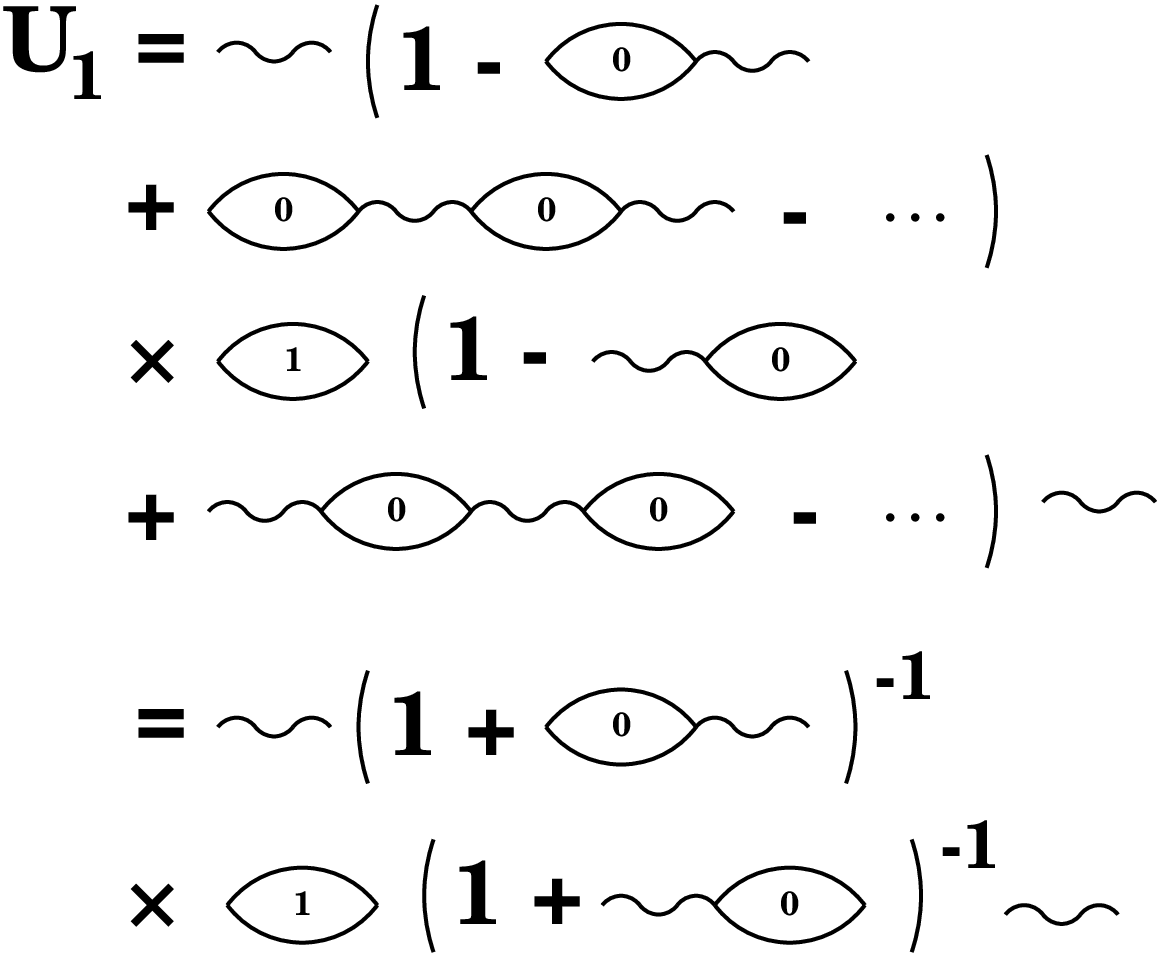,width=6.cm}}
\vspace{0.3cm}
\caption{RPA approximation for the two-particle potential $U$. The
potential $U_0$ is given by the same sequence of diagrams as the
potential $\Phi_0$ in Fig.~1 provided the external dashed line is
replaced by the Coulomb interaction (the wavy line).}
\label{fig2}
\end{figure}

\subsection{Weak localization effects}

The potential $U(\bbox{r}, \bbox{r'})$ can be found in the RPA
approximation \cite{BMM} (see Fig.~2). As
in the case of the polarizability, it is convenient to split the
polarization function into two parts (\ref{polar4}) and
(\ref{polar3}). Assuming the low frequency limit $\omega\ll\Delta$,
we replace the function $A(\omega)$ by its zero-frequency value,
Eq.(\ref{aw}). In the zeroth order in $\Pi_1$ one should solve an
equation
\begin{eqnarray} \label{RPA1}
U_0(\bbox{r}, \bbox{r'}) & = & V_0(\bbox{r} - \bbox{r'}) -
2\int_{\Omega} d\bbox{r_1} d\bbox{r_2} V_0(\bbox{r} - \bbox{r_1}) 
\nonumber \\
& \times & \Pi_0 (\bbox{r_1}, \bbox{r_2}) U_0 (\bbox{r_2}, \bbox{r'}),
\ \ \ V_0(r) = e^2/r. 
\end{eqnarray}
This equation was solved for an arbitrary closed system in
Ref. \cite{BMM}. The result is
\begin{equation} \label{RPAres}
U_0(\bbox{r}, \bbox{r'}) = \bar U + \tilde \Phi_0 (\bbox{r}) + \tilde 
\Phi_0 (\bbox{r'}) + U_{\kappa} (\bbox{r}, \bbox{r'}).
\end{equation}
Here $\bar U \equiv (e^2/C)_0$ is a constant, corresponding to the
charging energy calculated in the Thomas-Fermi approximation; 
$U_{\kappa}$ is the
usual screened Coulomb potential, shifted by a constant
 so that $\int d\bbox{r}
U_{\kappa} (\bbox{r}, \bbox{r'}) = 0$, while $\tilde \Phi_0$ is the
contribution due to the excess positive charge, moved towards the
boundary of the system after an extra electron is added to the
system. For the sphere (3D) and circle (2D) geometries this potential
has an explicit form \cite{BMM} 
\begin{eqnarray*} 
\tilde \Phi_0 (\bbox{r}) = \mbox{const}-\left\{ \begin{array}{lr}
e^2 (\kappa R^2)^{-1} \exp(-\kappa(R-r)), & 3D\\
e^2 (2\kappa R)^{-1} (R^2 - r^2)^{-1/2}, & 2D 
\end{array} \right., 
\end{eqnarray*}
where the constant is chosen in such a way that
 $\int d\bbox{r} \tilde \Phi_0 (\bbox{r}) =0$.

In the first order in $\Pi_1$ we obtain (Fig.~2)
\begin{eqnarray*}
U_1(\bbox{r}, \bbox{r'}) = 2\int d\bbox{r_1} d\bbox{r_2}
U_0(\bbox{r},\bbox{r_1}) \Pi_1 (\bbox{r_1},\bbox{r_2}) U_0
(\bbox{r_2}, \bbox{r'}),  
\end{eqnarray*}
and, taking into account that the integral of $\Pi_1$ over any of the
coordinates is zero, we write the corresponding contribution to
the charging energy (which constitutes the WL correction) in the form 
\begin{equation} \label{capWL}
\left( \frac{e^2}{C} \right)_1 = 2\int d\bbox{r_1} d\bbox{r_2}
\tilde \Phi_0(\bbox{r_1}) \Pi_1 (\bbox{r_1},\bbox{r_2}) \tilde \Phi_0 
(\bbox{r_2}).  
\end{equation}
Due to the structure of the function $\Pi_1$ (Eq. (\ref{polar1}))
there are two contributions to the WL correction: one comes from the
short-ranged term, and another one is related to the diffusion. A
comparison of these contributions can be carried out exactly in the
same way as was done for the polarizability, and it turns out that the
diffusive term dominates if the sample size exceeds considerably the
mean free path. Thus,  we obtain finally 
\begin{equation} \label{cap1}
\left( \frac{e^2}{C} \right)_1 = -\frac{4\nu }{\beta V} \int d\bbox{r_1}
d\bbox{r_2} \tilde \Phi_0(\bbox{r_1}) \Pi_D (\bbox{r_1},\bbox{r_2})
\tilde \Phi_0 (\bbox{r_2}).
\end{equation}
The calculation for the particular  geometries of a sphere (3D) and
disk (2D) gives
\begin{eqnarray} \label{cap2}
\left( \frac{e^2}{C} \right)_1 = \left\{ \begin{array}{lr} 
-1.32\beta^{-1}\tau^{-1}(p_F R)^{-4}, & \mbox{\rm 3D} \\
-0.010\beta^{-1} \tau^{-1} (p_FR)^{-2}, &
\mbox{\rm 2D} \end{array} \right. .
\end{eqnarray}
The small coefficient in front of the 2D expression is an
artifact of  the specific circle geometry. 

The weak localization correction suppresses the charging energy,
i.e. enhances the capacitance. The magnetic field suppresses the
capacitance. Both in 2D and 3D cases, the WL correction to the
charging energy can be estimated as  
\begin{equation} \label{cap3}
\left( \frac{e^2}{C} \right)_1 \sim \frac{\Delta}{g}.
\end{equation}
The WL correction can be in principle extracted from the measurements
of the magnetic field dependence of the capacitance, though its rather
small value may make such a measurement problematic. 

As for the polarizability, the CE calculation yields an additional factor
$-2.75$ in Eqs. (\ref{cap1}) and (\ref{cap2}). Thus, in CE the
magnetic field suppresses the charging energy, enhancing the
capacitance.    

\section{Mesoscopic fluctuations}

Mesoscopic fluctuations of the polarizability and the capacitance can
be calculated in a similar way. In addition to the average
polarization function (\ref{polar3a}), (\ref{polar3}) there is also a
random part $\Pi_r (\bbox{r}, \bbox{r'})$ with the zero average,
giving rise to the fluctuations of these quantities. Since the
integral of $\Pi_r$ over each coordinate is zero, we immediately
arrive to the expressions for the random parts of the static
\cite{foot1} polarizability $\alpha_r$ and the charging energy
$(e^2/C)_r$ in precisely the same way as Eqs. (\ref{dip1}) and
(\ref{capWL}) were obtained: 
\begin{eqnarray} \label{fl1p} 
\alpha_r & = & \frac{2e^2}{E^2} \int d\bbox{r} d\bbox{r'}
\Phi_0(\bbox{r}) 
\Pi_r (\bbox{r},\bbox{r'}) \Phi_0(\bbox{r'})
\end{eqnarray}
and
\begin{eqnarray} \label{fl1c}
\left( \frac{e^2}{C} \right)_r & = & 2\int d\bbox{r} d\bbox{r'}
\tilde \Phi_0 (\bbox{r}) 
\Pi_r (\bbox{r},\bbox{r'}) \tilde \Phi_0 (\bbox{r'}). 
\end{eqnarray}

\begin{figure}
\centerline{\psfig{figure=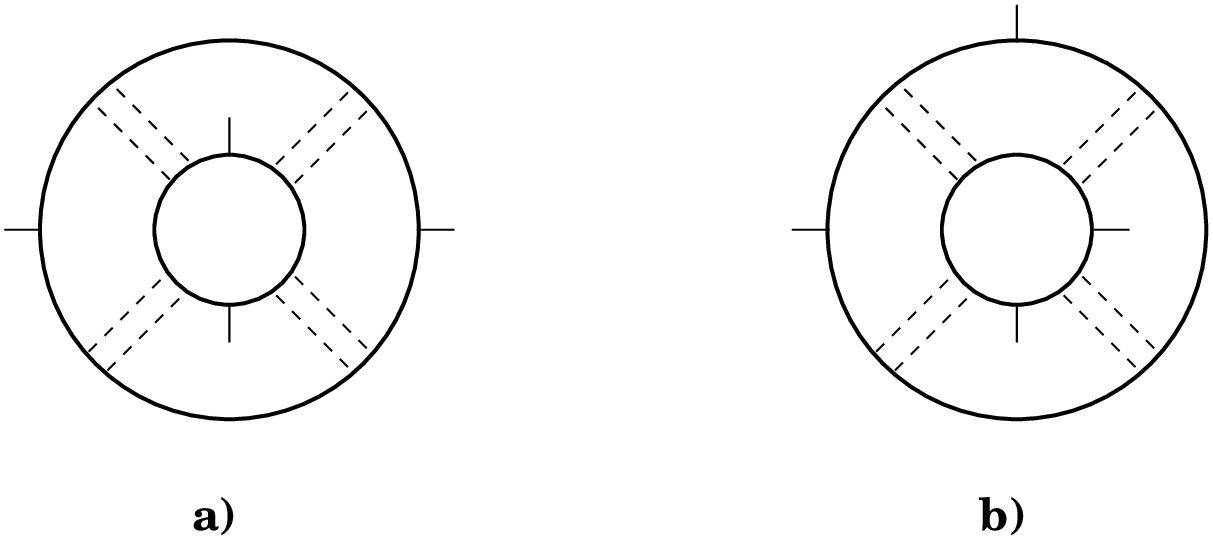,width=7.cm}}
\vspace{0.3cm}
\caption{Diagrams for the fluctuations of the polarization
function. The double dashed lines denote the diffusion
propagators. The counting factors are 2 (a) and 4 (b) for the unitary
symmetry, and 4 (a) and 8 (b) for the orthogonal one.}
\label{fig3}
\end{figure}

Thus,  the mesoscopic fluctuations of both
quantities are determined by the fluctuations of the polarizability. 
To calculate them, 
we perform the perturbative calculation. As we will see, in the case of
fluctuations the whole range of energies $\Delta \lesssim \epsilon
\lesssim E_c$ contributes, and this fact justifies the perturbative
calculation (in contrast to the weak localization correction, which is
determined by low energies). Following Berkovits and Altshuler
\cite{Berk}, we identify the four-diffusion diagrams (Fig.~3)  as
giving the main contribution to fluctuations of the polarization
function.  

We obtain 
\begin{eqnarray} \label{flpolar}
& & \langle \Pi_r (\bbox{r_1}, \bbox{r_3}) \Pi_r (\bbox{r_2},
\bbox{r_4}) \rangle = -\frac{(12/\beta)}{2\pi^2} \mbox{Re} \int_0^{\infty}
\epsilon d\epsilon D_{\epsilon} (\bbox{r_1}, \bbox{r_2}) \nonumber \\
& & \times D_{\epsilon} (\bbox{r_2}, \bbox{r_3}) D_{\epsilon}
(\bbox{r_3}, \bbox{r_4}) D_{\epsilon} (\bbox{r_4}, \bbox{r_1}). 
\end{eqnarray}
Here $12/\beta$ is a combinatorial factor (a number of 
4-diffuson diagrams); the function $D_\epsilon$ is given by the
following expression,
\begin{equation} \label{difl}
D_{\epsilon} (\bbox{r}, \bbox{r'}) = -\frac{1}{i\epsilon V} + \pi\nu
\Pi_D (\bbox{r}, \bbox{r'}), \ \ \ \Delta \lesssim \epsilon \lesssim
E_c \ ,
\end{equation}
and decreases for $\epsilon \ge E_c$. If the integral over $\epsilon$
in Eq. (\ref{flpolar}) diverges, the cut-off at energies of order of
the mean level spacing $\Delta$, where the perturbative expression
(\ref{flpolar}) ceases to be valid, should be introduced.  

A naive estimate suggests that
the leading contribution to the fluctuations of the polarizability as
well as of the capacitance can be found by substituting the 
 zero-mode contribution (the first
term in rhs of Eq. (\ref{difl}))  for all the four
functions $D_\epsilon$. However, since the integrals of the potentials
$\Phi_0$ and $\tilde \Phi_0$ are zero, this term 
vanishes. For typical geometries the main contribution
is given by the term  
where two of the functions $D_{\epsilon}$ are replaced by the
zero-mode result, while two others are represented by the diffusion
propagator (the second term in rhs of Eq. (\ref{difl})). Cutting off
the logarithmically divergent integral at $\Delta$ from below and at
the Thouless energy $E_c$ from above, we obtain 
\begin{eqnarray} \label{fl2p}
\langle \alpha_r^2 \rangle & = & \frac{12e^4}{\beta E^4} \nu^2 \ln g
\nonumber 
\\ & \times & \left[
\frac{2}{V} \int d\bbox{r_1} d\bbox{r_2} \Phi_0(\bbox{r_1}) \Pi_D
(\bbox{r_1},\bbox{r_2}) \Phi_0(\bbox{r_2}) \right]^2
\end{eqnarray}
and
\begin{eqnarray} \label{fl2c}
\langle \left( \frac{e^2}{C} \right)_r^2 \rangle & = & {12\over\beta}
\nu^2 \ln g \\ & \times & \left[
\frac{2}{V} \int d\bbox{r_1} d\bbox{r_2} \tilde \Phi_0 (\bbox{r_1})
\Pi_D (\bbox{r_1},\bbox{r_2}) \tilde \Phi_0 (\bbox{r_2})
\right]^2. \nonumber  
\end{eqnarray}
Comparing these results
with the expression for the WL corrections, 
Eqs. (\ref{diffpol}) and (\ref{cap1}), we obtain a general
formula relating WL corrections (calculated within GCE) to the
mesoscopic fluctuations of the same quantity,
\begin{equation} \label{fl3p}
\mbox{r.m.s} \ (\alpha) = ( 3\beta \ln g)^{1/2} 
\left\vert \alpha_1 \right\vert 
\end{equation}
and
\begin{equation} \label{fl3c}
\mbox{r.m.s} \left( \frac{e^2}{C} \right) = (3\beta \ln g)^{1/2} 
\left\vert \left( \frac{e^2}{C} \right)_1 \right\vert. 
\end{equation}
Note that in contrast to the WL corrections, the mesoscopic
fluctuations are not expected to be sensitive to the GCE/CE
difference, since the integral (\ref{flpolar}) is determined equally
by all energies $\Delta \lesssim \epsilon \lesssim E_c$. 

For particular geometries we obtain 
\begin{eqnarray*} 
\mbox{r.m.s} \ \alpha = \left( \frac{\ln g}{\beta} \right)^{1/2}
\left\{ \begin{array}{lr}  4.71 l^{-1}(p_F \kappa)^{-2}, & \mbox{\rm 3D} \\
5.30 R^2 (\kappa p_F l)^{-1}, &
\mbox{\rm 2D} \end{array} \right. 
\end{eqnarray*}
and
\begin{eqnarray*} 
\mbox{r.m.s} \ \frac{e^2}{C} = \left( \frac{\ln g}{\beta}
\right)^{1/2} \left\{ \begin{array}{lr}  2.29 \tau^{-1}(p_F R)^{-4}, &
\mbox{\rm 3D} \\ 
0.017 \tau^{-1} (p_FR)^{-2}, &
\mbox{\rm 2D} \end{array} \right. .
\end{eqnarray*}

The remaining terms in Eq. (\ref{flpolar}), with all the four
functions $D_{\epsilon}$  replaced by diffusion propagator, can
also be easily estimated. Their contribution is of the same order as
Eqs. (\ref{fl3p}), (\ref{fl3c}), but without the logarithmic factor in
the numerator. Thus, for the r.m.s. of the charging energy in
addition to the term of order $\sim \Delta (\ln g)^{1/2} /g$ (two
zero modes) we obtain a correction of order $\sim \Delta/g$ (no zero
modes).  

Amazingly, the relations between the weak localization correction and
the amplitude of mesoscopic fluctuations, 
Eqs. (\ref{fl3p}) and (\ref{fl3c}), have a universal form, the same
for 2D and 3D systems. We should note, however, that these results 
are not applicable for the case of polarizability fluctuations
of a quasi-two-dimensional sample of the thickness $h
\gg l$ and area $S \gg h^2$ in the transverse field (see Sec.\ref{s2b})  
calculated previously by Berkovits and Altshuler \cite{Berk}.
In this case, the contribution from the terms with two zero-modes,
Eq. (\ref{fl2p}), which can be easily calculated with the use of
Eq. (\ref{potq2d}), yields the r.m.s value of the polarizability
fluctuations,  
$$\mbox{r.m.s} (\alpha^{(2)}) = \frac{(3 \beta^{-1}\ln g)^{1/2}}{\pi g}
\frac{h}{\kappa^2},$$
where $g=2\pi\nu Dh$. 
On the other hand, the contribution where all the four functions
$D_\epsilon$ are  replaced by the diffusion propagators is \cite{Berk}
$$\mbox{r.m.s} (\alpha^{(4)}) =  \frac{8\beta^{-1}\pi^{3/2}}{g} 
\frac{S^{1/2}}{\kappa^2},$$   
i.e it is larger by a factor $\sim S^{1/2}/[h(\ln g)^{1/2}]$, 
and for this particular geometry represents the leading contribution
to the fluctuations of the polarizability.

\section{Conclusions} 

In this paper, we have calculated the weak localization (WL)
correction to the polarizability and the capacitance of a disordered
sample and the mesoscopic fluctuations of these quantities. The WL
corrections originate from the $G_RG_A$ term in the polarization
function, which depends on the presence or absence of the
time-reversal symmetry.  A change of the polarization function
influences the screening and, consequently, the polarizability and the
capacitance. We find that in the grand canonical ensemble, switching
on the magnetic field leads to a positive correction to the
polarizability and negative one to the capacitance. In the canonical
ensemble the magnitude of the effect is the same (up to a numerical
coefficient $\sim 2.75$), however the sign is reversed.  

Calculating the mesoscopic fluctuations of the polarizability, we find
that for typical geometries they are related to the value $\alpha_1$
of the WL correction as follows (see Eq. (\ref{fl3p})):
$$ \mbox{r.m.s.}(\alpha)=(3\beta\ln g)^{1/2}|\alpha_1|. $$
The same conclusion is valid for the capacitance, see Eq.(\ref{fl3c}).
Therefore, the magnitude of fluctuations exceeds considerably the
value of the WL correction. This should be contrasted with the relation
of the corresponding quantities for the case of the conductance:
\begin{eqnarray*}
&&\mbox{r.m.s.}(g)\sim 1;\\
&&|g_1|\sim\left\{
\begin{array}{ll}
1\ , & \qquad \mbox{quasi-}1D\\
\ln(L/l)\ , & \qquad 2D\\
L/l\ , & \qquad 3D\ ,
\end{array}
\right.
\end{eqnarray*}
so that $|g_1|\gg\mbox{r.m.s.}(g)$ in 2D and 3D and
$|g_1|\sim\mbox{r.m.s.}(g)$ in the quasi-1D geometry. 
As our results show, an experimental observation of the
magnetopolarizability of mesoscopic samples requires an experimental
setup with large number of such samples, which would reduce the
fluctuations. 

Mesoscopic fluctuations of the charging energy contribute to the
fluctuations of the conductance peak spacings in the addition spectra
of quantum dots in the Coulomb blockade regime
\cite{Marcus2,Sivan1,Wharam,Marcus3}. However, as follows from
Eqs.(\ref{cap3}), (\ref{fl3c}), the magnitude of these fluctuations
is much smaller (by a factor $\sim(\ln g)^{1/2}/g$) than the level
spacing $\Delta$. Therefore, the contribution of the charging energy
fluctuations to the fluctuations of the peak spacings is
parametrically smaller than the effect of electron level fluctuations,
which is given by the random matrix theory and is of order of
$\Delta$. The charging energy fluctuations represent one (but not the
only one) of the contributions to the enhancement of the peak spacing
fluctuations as compared to the random matrix theory. This problem was
considered in detail in Ref.\cite{BMM}.

Finally, we would like to mention once more that we assumed the
screening length to be much larger than the wave length, or in other
words, that $r_s\ll 1$, where $r_s=e^2/\epsilon v_F$. In the opposite
case, $r_s>1$ (but still  below the Wigner crystallization
threshold), one can roughly estimate the result assuming the screening
length to be approximately given by the distance between electrons. 
This leads to an enhancement of the above
results for the WL correction and
the r.m.s. amplitude of fluctuations by a factor of order of $r_s$
(resp. $r_s^2$) for the polarizability and capacitance, respectively.

\section*{Acknowledgments}

We are grateful to R.~Berkovits and H.~Bouchiat for discussions
and comments. The work was supported by the Swiss National Science
Foundation (Y.~M.~B.) and SFB 195 der Deutschen Forschungsgemeinschaft
(A.~D.~M.).   

\section*{Appendix A. Magnetopolarizability in the canonical ensemble}

In this Appendix we calculate the magnetopolarizability $\alpha_B$
defined by Eq. (\ref{magn}) in the canonical ensemble. For this
purpose we first rewrite our derivation of the Sec. II in terms of
exact eigenfunctions and energy eigenvalues, in the same manner it was
done originally by GE \cite{GE}, and later by NRB \cite{Noat}. Using
the results, derived for the correlation of eigenfunctions in
Refs. \cite{BM1,BM2} (unitary ensemble) and Appendix B (orthogonal
ensemble), we first obtain the weak localization correction in the
grand canonical ensemble. It is in full agreement with the results
obtained in Sec. II. Then we generalize the derivation to the CE case.

Our conclusion is that the CE magnetopolarizability, $\alpha_B^{CE}$,
can be easily obtained from the GCE value, $\alpha_B^{GCE}$, as
follows:
\begin{equation} \label{CEbas}
\alpha_B^{CE} = -A_{CE} \alpha_B^{GCE}, \ \ \ A_{CE} = 2.75.
\end{equation}
We should stress here that the derivation given below uses only the
linear response formalism and the properties of the eigenfunction and
eigenvalue statistics in disordered systems. It relies on the fact
that that the WL correction to the polarizability is determined by the
energy range where the level correlation is important. It can be
repeated for the case of the capacitance, where one obtains a relation
analogous to Eq. (\ref{CEbas}). In contrast to this, the fluctuations of
both quantities are determined by the energies $\Delta \ll \epsilon
\ll E_c$, where the correlation of levels does not play any role, and
the difference between CE and GCE values is not expected. 

\subsection*{Magnetopolarizability in GCE: derivation \`a la Gor'kov
and Eliashberg}

We start from Eq. (\ref{dip1}), which can be rewritten in terms of the
exact single-particle states as follows, 
\begin{eqnarray} \label{gen1}
\alpha_B (\omega) = \frac{2e^2}{E^2} \int d\bbox{r_1} d\bbox{r_2}
\Phi_0(\bbox{r_1}) \bbox{\delta} \Pi(\bbox{r_1}, \bbox{r_2}) \Phi_0
(\bbox{r_2}). 
\end{eqnarray}
Here we introduced the symbol $\bbox{\delta}$ denoting the difference
between the quantities in unitary and orthogonal ensembles,
$$\bbox{\delta} (\dots) = (\dots)_{GUE} - (\dots)_{GOE},$$
and the polarization function $\Pi$ is expressed as follows,
\begin{eqnarray} \label{gen2}
\Pi(\bbox{r_1}, \bbox{r_2}) & = & \sum_{m \ne n} \psi_m^*(\bbox{r_1})
\psi_n (\bbox{r_1}) \psi_n^*(\bbox{r_2}) \psi_m (\bbox{r_2}) \nonumber
\\
& \times & \frac{n_F(\epsilon_m) - n_F(\epsilon_n)}{\omega -
\epsilon_m + \epsilon_n + i0}, 
\end{eqnarray}
$m$ and $n$ being the exact single-particle states. Thus, for $\omega
\ll \Delta$ we obtain an
expression valid both in GCE and CE,
\begin{eqnarray} \label{gen3}
\alpha_B (\omega) = \frac{4e^2}{E^2} \bbox{\delta} \left\langle
\sum_{\epsilon_n < \epsilon_F < \epsilon_m} \frac{1}{\epsilon_m -
\epsilon_n} \left\vert \left( \Phi_0 \right)_{mn} \right\vert^2
\right\rangle.   
\end{eqnarray}

In GCE the position of the Fermi level can be arbitrary, and we
replace the sum in Eq. (\ref{gen3}) by an integral with the level
correlation function $R_2 (\epsilon)$:
\begin{equation} \label{GCElev}
\sum_{\epsilon_n < \epsilon_F < \epsilon_m} (\dots) = \Delta^{-2}
\int_0^{\infty} \epsilon d\epsilon R_2(\epsilon) (\dots)\ \ . 
\end{equation} 

Using the sum rule for the eigenfunctions, 
$$\left\langle \sum_n  \psi_m^*(\bbox{r_1}) \psi_n (\bbox{r_1})
\psi_n^*(\bbox{r_2}) \psi_m (\bbox{r_2}) \right\rangle = V^{-1}
\delta(\bbox{r_1} - \bbox{r_2}),$$ 
we obtain for the magnetopolarizability
\begin{equation} \label{GCE1}
\alpha_B = - \frac{2e^2}{E^2 \Delta} \bbox{\delta} \left\langle
\left\vert \left( \Phi_0 \right)_{mm} \right\vert^2 \right\rangle. 
\end{equation}

The above derivation of Eq. (\ref{GCE1}) is essentially equivalent to
that of Refs. \cite{GE,Noat}. Now, using Eq. (\ref{fin1}) for the case
of the orthogonal symmetry and Ref. \cite{BM2} for the unitary one, we
write ($r = \vert \bbox{r_1 - r_2} \vert$)
\begin{eqnarray*}
& & V^2\langle \vert \psi_m(\bbox{r_1}) \psi_m(\bbox{r_2}) \vert^2
\rangle_{\epsilon} \nonumber \\ 
& & = \left\{ \begin{array}{ll}
\left[ 1+ 2\Pi_D (\bbox{r_1}, \bbox{r_2})\right] \left[1 + 2k_d(r)
\right], & \ \ \ GOE \\
\left[ 1+ \Pi_D (\bbox{r_1}, \bbox{r_2})\right] \left[1 + k_d(r)
\right], & \ \ \ GUE \end{array} \right. .
\end{eqnarray*}
Separating the dominating  diffusion
term, we obtain for the magnetopolarizability in the grand canonical
ensemble 
\begin{equation} \label{fin22}
\alpha_B = \frac{2e^2}{V^2E^2\Delta} \int d\bbox{r_1} d\bbox{r_2}
\Phi_0 (\bbox{r_1}) \Pi_D (\bbox{r_1}, \bbox{r_2}) \Phi_0
(\bbox{r_2}),
\end{equation}  
which coincides with Eq. (\ref{diffpol}). 

\subsection*{Canonical ensemble}

To realize the canonical ensemble, we apply the method previously
developed in Refs. \cite{Altland,Schmid}. Namely, we fix the number
of electrons to be integer in each individual sample, but allow it to
fluctuate slightly from sample to sample. This type of ensemble is
realized by pinning the Fermi-level to one of the single-particle
levels $\epsilon_k$: $\epsilon_F = \epsilon_k + 0$. 

Now instead of Eq. (\ref{GCElev}), one should split the sum over
energy levels in Eq. (\ref{gen3}) into two. The first contribution
consists of the terms with $m=k$, and can be transformed to the
integral with the use of the two-level correlation function $R_2$. The
rest of the sum requires the three-level correlator $R_3(0, \epsilon,
\epsilon_1)$, which corresponds to the probability to find three levels
with energies $0, -\epsilon_1$, and $\epsilon-\epsilon_1$ (counted
from the Fermi-surface). Thus, we obtain
\begin{eqnarray} \label{CElev}
\sum_{\epsilon_n < \epsilon_F < \epsilon_m} (\dots) & = & \Delta^{-1} 
\int_0^{\infty} d\epsilon R_2(\epsilon) (\dots) \\
& + & \Delta^{-2}
\int_0^{\infty} d\epsilon \int_0^{\epsilon} d\epsilon_1 R_3(0,
\epsilon_1, \epsilon) (\dots)\ \ , \nonumber 
\end{eqnarray} 
and for the magnetopolarizability
\begin{eqnarray} \label{ce1}
\alpha_B & = & \frac{4e^2}{E^2\Delta^2} \bbox{\delta} \left\{
\int_0^{\infty} \frac{d\epsilon}{\epsilon} \right. \\
& \times & \left[ R_2(\epsilon) \Delta
 +   \left. \int_0^{\epsilon} d\epsilon_1 R_3(0,\epsilon,
\epsilon_1) \right] \left\langle \left\vert \left( \Phi_0 \right)_{mn}
\right\vert^2 \right\rangle_{\epsilon} \right\} . \nonumber
\end{eqnarray}
Since the integral over $\epsilon$ converges for $\epsilon \sim \omega$,
we can replace the matrix element by its low-frequency limit. For
$\epsilon \ll E_c$ we obtain for the correlation of the wavefunctions
with $m \ne n$ (see Eq. (\ref{states}) for the orthogonal symmetry and
Ref. \cite{BM2} for the unitary one): 
\begin{eqnarray*}
& & V^2\langle \psi^*_m(\bbox{r_1}) \psi_n(\bbox{r_1}) 
\psi_m(\bbox{r_2}) \psi^*_n(\bbox{r_2}) \rangle_{\epsilon_0, \epsilon}
\nonumber \\ 
& & = \left\{ \begin{array}{ll}
k_d(r) + \left[1 + k_d(r) \right] \Pi_D (\bbox{r_1}, \bbox{r_2}), &\ \
\ GOE \\
k_d(r) + \Pi_D (\bbox{r_1}, \bbox{r_2}), &\ \ \ GUE 
\end{array} \right. . 
\end{eqnarray*}
In the cases when the diffusion dominates \cite{foot2}, one can
replace this eigenfunction correlator for either symmetry by
$\Pi_D$. As a result, taking into account Eq. (\ref{fin22}), we obtain
Eq. (\ref{CEbas}), with the coefficient $A_{CE}$ expressed in terms of
the level correlation functions, 
\begin{eqnarray} \label{coefa}
A_{CE} & = & -1 - 2\int_0^{\infty} \frac{ds}{s} \\
& \times & \bbox{\delta} \left[ R_2(s)  
 +   \int_0^s ds_1 \left( R_3(0,s_1,s) - R_2(s) \right)
\right], \nonumber 
\end{eqnarray} 
where we have made a change of variables $s = \pi
\epsilon/\Delta$. Note that Eq. (\ref{coefa}) differs from the similar
expression derived by NRB \cite{Noat} by the second term in the rhs.  

In the leading approximation the correlation function $R_2$ and $R_3$ may
be taken from the random matrix theory \cite{Mehta}. The first
integral in the rhs of Eq. (\ref{coefa}) can be calculated
analytically and is given by
\begin{eqnarray*}
I_1 & =& \int_0^{\infty} \frac{ds}{s} \bbox{\delta} R_2(s) =
\int_0^{\infty} \frac{ds}{s} g(s) h(s) = \frac{1}{4} -
\frac{\pi^2}{16} \\ & \simeq & -0.367,  
\end{eqnarray*}
in contrast to the statement of NRB that it is equal to $-1/2$. We
have defined the functions 
$$f(s) = \frac{\sin s}{s}; \ \ g(s) = \frac{d}{ds} f(s); \ \ h(s) =
\int_s^{\infty} f(s_1) ds_1.$$ 
The second term in the rhs of Eq. (\ref{coefa}) after a lengthy
algebra can be expressed as follows:
\begin{eqnarray*}
& &I_2= \int_0^{\infty} \frac{ds}{s} \int_0^s ds_1 \bbox{\delta} \left(
R_3(0,s_1,s) - R_2(s) \right) \\
& & = 2\int_0^{\infty} \frac{ds}{s}
\int_0^s ds_1 \left\{ g(s_1) h(s_1) - f(s) g(s_1) h(s-s_1) \right. \\ 
& & \left. + g(s) f(s_1) h(s-s_1) + h(s) f(s_1) g(s-s_1) \right\}.
\end{eqnarray*}
Calculating this numerically, we find $I_2=-1.509$ and thus, 
$A_{CE} = -1-2(I_1+I_2)=2.753$. 

Note that in the above derivation we neglected the contribution of the
so-called Debye processes (relaxation to the instantaneous equilibrium
distribution due to coupling with phonons or other possible inelastic
processes) \cite{Land,Gefen,Noat}. These processes do not exist in a
closed sample  in the limit of zero temperature ($T\ll\Delta$)
that we are considering  \cite{debye}. 

\section*{Appendix B. Correlations of eigenfunctions in disordered
systems: orthogonal ensemble}

In this Appendix we derive the expressions for the correlations of the
eigenfunctions in the orthogonal ensemble in the same way as these
were obtained in Ref. \cite{BM2} for the unitary ensemble. We restrict
ourselves to the terms of order $g^{-1}$.

Following \cite{BM2} we define the eigenfunctions correlators (see
Eq. (\ref{aver0})),
\begin{eqnarray} \label{fun}
& & \eta(\bbox{r_1}, \bbox{r_2}, \epsilon) = \left\langle
\vert \psi_k(\bbox{r_1}) \psi_k(\bbox{r_2}) \vert^2
\right\rangle_{\epsilon} \nonumber \\
& & \qquad \equiv \displaystyle{\frac{\langle \sum_k \vert
\psi_k(\bbox{r_1}) \psi_k(\bbox{r_2}) \vert^2 \delta (\epsilon -
\epsilon_k) \rangle}{\langle \sum_k \delta (\epsilon - \epsilon_k)
\rangle}}, \\  
& & \beta(\bbox{r_1}, \bbox{r_2}, \epsilon, \omega) =
\left\langle \vert \psi_k(\bbox{r_1}) \psi_l (\bbox{r_2})
\vert^2 \right\rangle_{\epsilon,\omega}, \ \ \ k \ne l, \nonumber 
\end{eqnarray}
and
\begin{eqnarray*} 
& & \gamma(\bbox{r_1}, \bbox{r_2}, \epsilon, \omega) =
\left\langle \psi^*_k(\bbox{r_1}) \psi_l(\bbox{r_1}) 
\psi_k(\bbox{r_2}) \psi^*_l(\bbox{r_2}) \right\rangle_{\epsilon,
\omega} , \ \ k \ne l.
\end{eqnarray*}
The quantities $\eta$ and $\beta$ are related as follows,
\begin{eqnarray} \label{a3}
& & B \equiv \eta (\bbox{r_1},\bbox{r_2}, \epsilon) \Delta^{-1}
\delta(\omega) + \beta(\bbox{r_1},\bbox{r_2},\epsilon, \omega)
\Delta^{-2} R_2(\omega) \nonumber \\
& & = \nu^2 + (2\pi^2)^{-1} \mbox{Re}
\left\{ \langle G^R(\bbox{r_1}, \bbox{r_1}, \epsilon) G^A
(\bbox{r_2}, \bbox{r_2}, \epsilon + \omega) \rangle \right.
\nonumber \\
& & - \left. \langle G^R(\bbox{r_1}, \bbox{r_1}, \epsilon) \rangle
\langle G^A (\bbox{r_2}, \bbox{r_2}, \epsilon + \omega)
\rangle \right\};
\end{eqnarray}
here the two-level correlation function, 
\begin{equation} \label{corf}
R_2(\omega) = \Delta^2 \left\langle \sum_{k \ne l} \delta(\epsilon -
\epsilon_k) \delta(\epsilon + \omega - \epsilon_l) \right\rangle ,
\end{equation}
is introduced.

The right-hand side of the expression (\ref{a3}) can
be directly calculated with the use of the supersymmetry
technique. For the case of preserved time-reversal symmetry
(orthogonal ensemble) one obtains 
\begin{eqnarray} \label{Q1}
& & B (\bbox{r_1}, \bbox{r_2}, \epsilon, \omega) = - (2\pi^2)^{-1}
\mbox{Re} \left\{ \left\langle g_{b1,b1}^{11} (\bbox{r_1}, \bbox{r_1})
\right. \right. \nonumber \\
& & \times \left. \left. g_{b1,b1}^{22} (\bbox{r_2}, \bbox{r_2})   
+ g_{b1,b1}^{12} (\bbox{r_1}, \bbox{r_2})
g_{b1,b1}^{21} (\bbox{r_2}, \bbox{r_1}) \right\rangle_F  
\right. \nonumber \\
& & - \left. \left\langle g_{b1,b1}^{11} (\bbox{r_1}, \bbox{r_1})
\right\rangle_F \left\langle g_{b1,b1}^{22} (\bbox{r_2},
\bbox{r_2}) \right\rangle_F \right\}.
\end{eqnarray}
Here $\langle \dots \rangle_F$ denotes the averaging with the action
of the supermatrix sigma-model $F[Q]$: 
\begin{eqnarray} \label{aver1}
& &\langle \dots \rangle_F = \int DQ ( \dots ) \exp(-F[Q]), \nonumber \\ 
& & F[Q] = - \frac{\pi\nu}{8} \int d\bbox{ r} \ \mbox{Str} [D(\nabla
Q)^2 + 2i(\omega + i0) \Lambda Q],
\end{eqnarray}
where $D$ is the diffusion coefficient, 
$Q = T^{-1}\Lambda T$ is a 8$\times$8 supermatrix, $\Lambda =
\mbox{diag} (1,1,1,1,-1,-1,-1,-1)$, and $T$ belongs to the supercoset space
$U(2,2 \vert 4)/U(2\vert 2)\times U(2 \vert 2)$. The symbol
$\mbox{Str}$ denotes the supertrace (trace over bosonic degrees of
freedom minus that over fermionic ones). The 
upper matrix indices correspond to the retarded-advanced
decomposition, while the lower indices denote the boson-fermion
one (here we need only the index $b1$, which denotes one of two
bosonic components of a supervector). 
The Green's function $g$ in Eq. (\ref{Q1}) is the solution to
the matrix equation:
\begin{eqnarray} \label{Green}
& & \left[ -i(\epsilon + \frac{\omega}{2} - \hat H_0) -
\frac{i}{2}(\omega + i0)\Lambda + Q/2\tau \right] g(\bbox{
r}, \bbox{ r}') \nonumber \\
& & = \delta (\bbox{ r} - \bbox{ r}').
\end{eqnarray}
Expressing these functions through the matrices $Q$ and taking into
account Eq. (\ref{a3}), we arrive at the following equation valid in
for an arbitrary diffusive system:
\begin{eqnarray} \label{genex}
& & 2\pi^2 \left[ \frac{\eta(\bbox{r_1}, \bbox{r_2},
\epsilon)}{\Delta} \delta(\omega) + 
\frac{\beta(\bbox{r_1}, \bbox{r_2},\epsilon, \omega)}{\Delta^2}  
R_2(\omega) \right] \nonumber \\
& & = (\pi\nu)^2 \mbox{Re} \left\{ 1 - \langle  Q_{b1,b1}^{11}
(\bbox{r_1}) Q_{b1,b1}^{22} (\bbox{r_2}) \right. \rangle_F \nonumber
\\  
& & \left. - 2k_d(r) \langle Q^{12}_{b1,b1} (\bbox{r_1})
Q^{21}_{b1,b1}(\bbox{r_1}) \rangle_F \right\}, 
\end{eqnarray}
here the function $k_d$ is defined in Eq. (\ref{kd}), and $r = \vert
\bbox{r_1} - \bbox{r_2} \vert$. The separation of the the rhs of
Eq. (\ref{genex}) into the singular (proportional to $\delta(\omega)$)
and regular parts allows one to obtain the quantities
$\alpha(\bbox{r}_1,\bbox{r}_2)$ and
$\beta(\bbox{r}_1,\bbox{r}_2,\omega)$.  

For the case of a metallic system in the weak localization
regime, the sigma-model correlation functions $\langle
Q_{b1,b1}^{11}(\bbox{r_1}) Q_{b1,b1}^{22} (\bbox{r_2}) \rangle_F$ and
$\langle Q_{b1,b1}^{12}(\bbox{r_1}) Q_{b1,b1}^{21} (\bbox{r_2})
\rangle_F$ can be calculated for relatively low frequencies $\omega
\ll E_c$ with the use of a general method developed in
Refs. \cite{KM,MF} which allows one to take into account spatial
variations of the field $Q$. The results are obtained in the form of
an expansion in $g^{-1}$. Up to the terms of order $g^{-1}$, we obtain
\begin{eqnarray*} 
& & \langle Q_{b1,b1}^{11}(\bbox{ r}_1) Q_{b1,b1}^{22} (\bbox{ r}_2)
\rangle_F \\
& & \qquad = 1 - 2\tilde R(\omega) - \frac{4i\Delta}{\pi (\omega +i0)}
\Pi_D(\bbox{r_1}, \bbox{r_2})   
\end{eqnarray*}
and
\begin{eqnarray*} 
& & \langle Q_{b1,b1}^{12} (\bbox{r_1}) Q_{b1,b1}^{21} (\bbox{r_2})
\rangle_F =  -2 \frac{i\Delta}{\pi (\omega +i0} \\
& & \qquad - 2 \left( \tilde
R(\omega) + \frac{i\Delta}{\pi (\omega + i0)} \right) \Pi_D(\bbox{r_1},
\bbox{r_2}).  
\end{eqnarray*}
Here the diffusion propagator $\Pi_D$ is defined by Eq. (\ref{diff}),
and we have introduced the function $\tilde R(\omega) = [1 + S(\omega)]/2$,
where $S(\omega)$ is given by Eq. (\ref{ort2}). Note that the two-level
correlation function, $R_2(\omega)$, is the real part of $\tilde
R(\omega)$.  
 
Now, separating regular and singular parts in rhs of Eq. (\ref{genex}),
we obtain the following result for the autocorrelations of the same 
eigenfunction,
\begin{eqnarray} \label{fin1}
& & V^2\langle \vert \psi_k(\bbox{r_1}) \psi_k(\bbox{r_2}) \vert^2
\rangle_{\epsilon} \nonumber \\
& & = \left[ 1 + 2k_d (r) \right] \left[ 1 + 2 \Pi_D (\bbox{r_1},
\bbox{r_2}) \right],
\end{eqnarray}
and for the correlation of amplitudes of two different eigenfunctions
($k \ne l$)
\begin{equation} \label{fin2}
V^2\langle \vert \psi_k(\bbox{r_1}) \psi_l(\bbox{r_2}) \vert^2
\rangle_{\epsilon, \omega} - 1  =  2 k_d(r) \Pi_D (\bbox{r_1},
\bbox{r_2}) . 
\end{equation}
The result (\ref{fin1}) for $\bbox{ r}_1 = \bbox{ r}_2$ is the inverse
participation ratio previously obtained in Ref. \cite{MF}, while that
for an arbitrary spatial separation was found in the zero-mode
approximation ($g = \infty$) in Ref. \cite{Prig}.    

Now we turn to the correlation function $\gamma$. Similarly to
Ref. \cite{BM2}, we obtain a relation 
\begin{eqnarray} \label{genex1}
& & 2\pi^2 \left[ \frac{\eta(\bbox{r_1}, \bbox{r_2},
\epsilon)}{\Delta} \delta(\omega) + 
\frac{\gamma(\bbox{r_1}, \bbox{r_2}, \epsilon, \omega)}{\Delta^2}  
R_2(\omega) \right] \nonumber \\
& & = - (\pi\nu)^2 \mbox{Re} \left\{ \langle  Q_{b1,b1}^{12}
(\bbox{r_1}) Q_{b1,b1}^{21} (\bbox{r_2}) \rangle_F + k_d(r)
\right. \\
& & \left. \times \left[ \langle Q^{11}_{bb} (\bbox{r_1})
Q^{22}_{bb}(\bbox{r_1}) \rangle_F + \langle
Q_{b1,b1}^{12}(\bbox{r_1}) Q_{b1,b1}^{21} (\bbox{r_2}) \rangle_F -1
\right] \right\}. \nonumber 
\end{eqnarray}
Separating again the rhs into the regular and singular parts, we recover
Eq. (\ref{fin1}) and obtain
\begin{eqnarray} \label{states}
& & V^2 \langle \psi^*_k(\bbox{r_1}) \psi_l(\bbox{r_1}) 
\psi_k(\bbox{r_2}) \psi^*_l(\bbox{r_2}) \rangle_{\epsilon, \omega}
\nonumber \\ 
& & \qquad = k_d(r) + \left[ 1 + k_d(r) \right]
\Pi_D(\bbox{r_1},\bbox{r_2}),\ \ \ k \ne l.  
\end{eqnarray}


\begin{thebibliography}{99}
\bibitem{foot3} Eq.(\ref{func1}) is written for a system with broken
time reversal symmetry (unitary ensemble); for the case of the
orthogonal ensemble see Appendix B, Eq.(\ref{states}).
\bibitem{foot1} In contrast to the weak localization correction, the
fluctuations of the polarizability $\langle \alpha_r^2 (\omega)
\rangle$ do not depend essentially on $\omega$ for $\omega<E_c$.
\bibitem{foot2} In the case when short-range terms dominate, the same
consideration holds, with $\Pi_D$ replaced  by $k_d$. Thus, 
the results  (\ref{coefa}), (\ref{CEbas}) for the coefficient $A_{CE}$
retain their validity also in this situation.
\end{thebibliography}
\end{document}